\documentclass[nofootinbib,preprint,superscriptaddress,showpacs,amsmath,amssymb,aps,pra,showkeys]{revtex4-1}

\usepackage{amsmath,amsfonts,amssymb}
\usepackage{graphicx}

\DeclareMathOperator{\arccot}{arccot}

\begin{document}

\title{Effects of the Coulomb potential in interference patterns of strong-field holography with photoelectrons}

\author{N. I. Shvetsov-Shilovski}
\email{n79@narod.ru}
\affiliation{Institut f\"{u}r Theoretische Physik and Centre for Quantum Engineering and Space-Time Research, Leibniz Universit\"{a}t Hannover, D-30167 Hannover, Germany}

\author{M. Lein}
\affiliation{Institut f\"{u}r Theoretische Physik and Centre for Quantum Engineering and Space-Time Research, Leibniz Universit\"{a}t Hannover, D-30167 Hannover, Germany}

\date{\today}

\begin{abstract}
Using the semiclassical two-step model for strong-field ionization we investigate the interference structures emerging in strong-field photoelectron holography, taking into account the Coulomb potential of the atomic core. For every kind of the interference pattern predicted by the three-step model, we calculate the corresponding structure in the presence of the Coulomb field, showing that the Coulomb potential modifies the interference patterns significantly.\\

 
\end{abstract}


\maketitle

\section{Introduction}

Development of techniques capable of tracing molecular dynamics involves fundamental and technological problems of great complexity that need to be solved. The reason is that the dynamic imaging techniques are to operate at sub-angstrom spatial scales with femtosecond time resolution. 
The continuous progress in laser technologies, especially the advent of the technology of pulse compression, as well as the advances in the development of the free electron lasers, have given rise to a variety of techniques aimed at time-resolved molecular imaging. Among these are: optical pump-probe spectroscopy, time-resolved electron and x-ray diffraction, and ultrafast x-ray spectroscopy (see Ref.~\cite{Agostini2016} for recent review).   

During the last three decades a new breakthrough in laser technology has been achieved: table-top intense femtosecond laser systems operating at various wavelenghts have become available in many laboratories all over the world. This has led to the emergence of such fields of research as strong-field, ultrafast, and attosecond physics (see Ref.~\cite{Krausz2009} for review). It was found that the interaction of intense laser radiation with atoms and molecules leads to a plethora of highly nonlinear phenomena. Among these are: above-threshold ionization (ATI) and the formation of the high-energy plateau in the electron energy spectrum (high-order ATI), generation of high-order harmonics (HHG), nonsequential double ionization (NSDI), etc (see Refs.~\cite{BeckerRev2002, MilosevicRev2003, FaisalRev2005, FariaRev2011} for recent reviews). The main theoretical approaches used in strong-field physics are the direct numerical solution of the time-dependent Schr\"{o}dinger equation (TDSE) (see, e.g., Refs.~\cite{Muller1999, Bauer2006, Madsen2007} and references therein), the strong-field approximation (SFA) \cite{Keldysh1964, Faisal1973, Reiss1980}, and semiclassical models using classical description of the electron after it has been promoted to the continuum, typically by tunneling ionization~\cite{Dau3, PPT, ADK}. The widely known semiclassical approaches are the two-step \cite{Linden1988, Gallagher1988, Corkum1989} and the three-step models \cite{Kulander_Schafer, Corkum1993}. 

The studies of ATI have shown that the vast majority of electrons reach the detector without recolliding with their parent ions. These electrons are referred to as direct ones and they have energies below $2U_{p}$, where $U_p=F^2/4\omega^2$ is the ponderomotive energy (atomic units are used throughout the paper unless indicated otherwise). There are also electrons that are driven back to their parent ions and scatter off them by angles close to 180$^{\circ}$. The high-energy plateau in the ATI spectrum is created due to these rescattered electrons. The rescattering scenario led to an understanding of the physical mechanisms responsible for HHG and NSDI. The returning electron can recombine with the parent ion and emit high-order harmonics \cite{Corkum1993, Krause1992}. As an alternative, if this electron has enough energy, it can release another electron from the atomic ion (see Ref.~\cite{Dorner2005} for review). These rescattering-induced processes can be qualitatively described within the three-step model. In the first step of this model an electron tunnels out of an atom, and in the second step it moves along a classical trajectory in the laser field only. The third step involves the interaction of the returning electron with the parent ion.  

Some of the phenomena mentioned here may be used for the development of new ways of time-resolved molecular imaging. Indeed, new ultrafast laser-based imaging techniques have been proposed recently: laser-assisted electron diffraction \cite{Kanya2010, Morimoto2014}, laser-induced Coulomb explosion imaging \cite{Frasinski1987, Cornaggia1991, Posthumus1995}, laser induced electron diffraction \cite{Lein2002, Meckel2008, Blaga2012, Pullen2015}, high-order harmonic orbital tomography \cite{Itatani2004, Haessler2010}, and strong-field photoelectron holography (SFPH) \cite{Huismans2011}.
 
Using the full coherence of the electron motion after ionization, the SFPH method puts into practice the idea of holography \cite{Gabor}. It was shown in Ref.~\cite{Huismans2011} that a photoelectron holographic pattern can be clearly recorded in experiment. The hologram is created by the interference between a reference (direct) electron and a signal (rescattered) one. The SFPH method of molecular imaging has several important advantages. First, although free-electron lasers were used in some of the SFPH experiments (see Refs~\cite{Huismans2011, Huismans2012}), this method can be realized in a table-top experiment. Second, the hologram that is recorded in SFPH encodes temporal and spatial information not only about the ion, but about the recolliding electron as well. Last but not least, attosecond time resolution can be achieved for the photoelectron dynamics. Indeed, the signal and reference electronic wave packets that produce the holographic patterns can be ionized in the same quarter cycle of the oscillating laser field. As a result, subcycle time resolution can be achieved even for long laser pulses.          

For these reasons, the SFPH has been studied extensively in the last few years, both experimentally (see, e.g., Refs.~\cite{Marchenko2011, Hickstein2012, Meckel2014, Haertelt2016, Walt2017}) and theoretically~\cite{Huismans2011, Huismans2012, Marchenko2011, Hickstein2012, Bian2011, Bian2012, Bian2014, Li2014, Li2015, Haertelt2016, Faria2017a, Faria2017b}. Among the theoretical approaches used to analyze the holographic structures are: the semiclassical model that accounts for the laser field only \cite{Bian2011, Bian2012, Bian2014, Li2014, Haertelt2016}, direct numerical solution of the TDSE \cite{Huismans2011, Marchenko2011, Hickstein2012, Bian2011, Haertelt2016}, the modified version of the SFA that accounts for the rescattering \cite{Huismans2011, Huismans2012}, the Coulomb-corrected strong-field approximation (CCSFA) \cite{Huismans2011, Huismans2012}, and the Coulomb quantum orbit strong-field approximation (CQSFA) \cite{Faria2017a, Faria2017b} (see Ref.~\cite{Faria2015} for the foundations of the CQSFA method). The holograms obtained from the solution of the TDSE agree with the experimental data. This is particularly true for the spacing between the side lobes (fringes) of the holographic structure emerging when both the signal and reference electrons are generated on the same quarter of cycle. However, it is difficult to interpret the hologram from the solution of the TDSE. 

The three-step semiclassical model was adapted for calculation of the SFPH, see Refs.~\cite{Bian2011, Bian2012, Bian2014, Li2014}. Different types of subcycle interference structures were predicted by this model~\cite{Bian2011}. Indeed, while the first studies of the SFPH considered only the interference of the reference and signal waves that are born in the same quarter cycle of the laser field, the signal and reference electrons can also originate from different quarter cycles, leading to different holographic patterns. Despite the appealing physical picture of the SFPH provided by the three-step model, it is well known that neglecting the Coulomb potential is severe (see, e.g., Refs.~\cite{Bashkansky1988, Paulus2000, Shvetsov2004}). Note that the same is true for the modified version of the SFA. Furthermore, the simulations of the SFPH within the three-step model were performed assuming that the starting point of the classical trajectory is independent on the field strength \cite{Bian2011, Bian2012, Bian2014, Li2014}. In contrast to this, it is natural to assume that electrons tunnel through a potential barrier with time-dependent width due to the oscillations of the laser field. It is known that the proper choice of the initial conditions for classical trajectories is important~\cite{Keller2012, Shvetsov2012}. 

The study of the electron trajectories calculated within the CCSFA showed that the trajectories responsible for the emergence of the interference structure can indeed be considered as reference and scattered wave packets \cite{Huismans2011}. The CCSFA approach reproduces the shape of the interference fringes and the spacing between them. Finally, the CCSFA simulations have given important insight into the role of the Coulomb potential in the SFPH~\cite{Huismans2011, Huismans2012}. The same is also true for the CQSFA theory that has allowed to identify and isolate many types of interference patterns in the photoelectron momentum distributions. The distortion of different kinds of interfering trajectories along with the change of the phase difference between them due to the presence of the ionic potential was studied in Ref.~\cite{Faria2017b}. However, no direct comparison of the interference structures predicted by the three-step model with those calculated taking the Coulomb potential into account has been made so far. 

Moreover, not all the principal types of interference structures predicted by the three-step model were considered in Refs.~\cite{Huismans2011, Huismans2012} and \cite{Faria2017b}. To the best of our knowledge, the effects of the Coulomb potential in the interference structures emerging due to the hard collisions of the signal electron with the atomic core have not been analyzed so far. Recall that hard collisions require small impact parameters and result in large changes of electron momenta including backward scattering. The interference of direct with \textit{backscattered} electrons has been proposed in Ref.~\cite{Bian2012} as a particularly sensitive probe of the molecular structure. The holograms for H$_2$ and N$_2$ measured recently in Ref.~\cite{Haertelt2016} revealed a fishbonelike structure that was claimed to originate from backward scattering. 

In this paper we revisit the holographic interferences calculated using the three-step model, in order to (i) understand how the time-dependent exit point affects the interference patterns, and (ii) to obtain a benchmark for comparison with the case when the Coulomb field is taken into account. We then calculate all major types of interference structures in the presence of the Coulomb potential, including those that involve hard collisions of the signal electron (backward scattering). Our analysis is based on the semiclassical two-step model (SCTS) that describes quantum interference and accounts for the Coulomb potential beyond the semiclassical perturbation theory (see Ref.~\cite{Shvetsov2016}). 

The paper is organized as follows. In Sec.~II we discuss the three-step model and its application to the SFPH when the starting point of the classical trajectories is time-independent, and when it depends on time. We formulate our approach to calculation of the SFPH with the Coulomb field in Sec.~III. In Sec.~IV we analyze formation of the interference structures in the presence of the Coulomb potential. The conclusions of the paper are given in Sec.~V.

\section{Strong-field photoelectron holography in the three-step model} 

\subsection{Application of the three-step model to strong-field photoelectron hologrpahy}

The application of the three-step model to the SFPH is reported in Refs.~\cite{Bian2011,Bian2012, Bian2014, Li2014, Li2015}. Here we repeat the main points that are important for the following discussion. For simplicity, and in order to be consistent with Refs.~\cite{Bian2011, Bian2012, Bian2014}, in this section we consider only one cycle of a linearly polarized cosine-like laser field: $\vec{F}\left(t\right)=F_0\cos\left(\omega t\right)\vec{e}_x$ between $\omega t=0$ and $\omega t=2\pi$. Here $\vec{e}_{x}$ is the unit vector in the polarization direction, $F_0$ is the field strength, and $\omega$ is the frequency. Newton's equation of motion for an electron moving in this field can be easily solved analytically. The velocity $\vec{v}\left(t\right)$ and the position $\vec{r}\left(t\right)$ of an electron launched at time $t_0$ are given by
\begin{equation}
\vec{v}\left(t\right)=\left\{v_x\left(t_0\right)+p_F\sin\left(\omega t_{0}\right)-p_F\sin\left(\omega t\right)\right\}\vec{e}_{x}+v_y\left(t_0\right)\vec{e}_{y}+v_z\left(t_0\right)\vec{e}_{z},
\label{velocity}
\end{equation}   
and 
\begin{eqnarray}
\vec{r}\left(t\right)& = & \left\{x_0\left(t\right)+\frac{F_0}{\omega^2}\left(\omega t-\omega t_{0}\right)\sin\left(\omega t_0\right)+\frac{F_0}{\omega^2}\left[\cos\left(\omega t\right)-\cos\left(\omega t_0\right)\right]\right\}\vec{e}_{x} \nonumber \\
& + & v_{y}\left(t_0\right)\left(t-t_0\right)\vec{e}_{y}+v_{z}\left(t_0\right)\left(t-t_0\right)\vec{e}_{z}. 
\label{position}
\end{eqnarray}
Here $\vec{v}\left(t_0\right)=v_x\left(t_0\right)\vec{e}_x+v_y\left(t_0\right)\vec{e}_y+v_z\left(t_0\right)\vec{e}_z$, $\vec{r}\left(t_0\right)=x\left(t_0\right)\vec{e}_x+y\left(t_0\right)\vec{e}_y+z\left(t_0\right)\vec{e}_z$, and $p_F=F_0/\omega$. Due to the cylindrical symmetry with respect to the polarization direction, we leave out the $z$-component of both $\vec{r}\left(t\right)$ and $\vec{v}\left(t\right)$ in what follows and consider electron motion in two spatial dimensions. 

We assume that the electron starts with zero initial velocity along the laser field $v_x\left(t_0\right)=0$, but its initial transverse velocity $v_{y}\left(t_0\right)\equiv v_{0,\perp}$ can be arbitrary. An electron starting with zero initial transverse velocity at a time instant $t_{0}^{sig}$ within a certain fraction of the laser period can return to the parent ion [i.e., to the point $\left(x=0, y=0\right)$]. Upon its return this signal electron is elastically scattered from the atomic core by an angle $\theta_{0}$ (see, e.g., Ref.~\cite{Paulus1994}). The scattering event is assumed to be instantaneous, and the scattering angle $\theta_{0}$ is randomly distributed between $0^{\circ}$ and $360^{\circ}$. 

The time of return $t_1$ of the signal electron can be found from the equation 
\begin{equation}
x\left(t_0\right)+\frac{F_0}{\omega^2}\left(\omega t_1-\omega t_0\right)\sin\left(\omega t_0\right)+\frac{F_0}{\omega^2}\left[\cos\left(\omega t_1\right)-\cos\left(\omega t_0\right)\right]=0.
\label{retcond}
\end{equation}
Since the signal electron returns to the core with the velocity 
\begin{equation}
V_{1}=-p_{F}\left[\sin\left(\omega t_1\right)-\sin\left(\omega t^{sig}_{0}\right)\right],
\label{vel_up_ret}
\end{equation}
its velocity at time $t>t_1$ is given by:
\begin{equation}
\vec{V}\left(t\right)=\left[V_{1}\cos\theta_{0}+p_{F}\sin\left(\omega t_{1}\right)-p_{F}\sin\left(\omega t\right)\right]\vec{e}_{x}+V_{1}\sin\theta_{0}\vec{e}_{y}.
\label{vel_after_sig}
\end{equation}
From Eq.~(\ref{vel_after_sig}) we obtain the asymptotic (final) momentum of the signal electron:
\begin{equation}
\vec{p}=\left[V_{1}\cos\theta_{0}+p_{F}\sin\left(\omega t_1\right)\right]\vec{e}_{x}+V_{1}\sin\theta_{0}\vec{e}_{y}.
\label{ass_mom_sig}
\end{equation}
The asymptotic momentum of a reference electron starting at time $t_0$ with initial transverse velocity $v_{0,\perp}$ reads as:
\begin{equation}
\vec{p}=p_{F}\sin\left(\omega t_0\right)\vec{e}_{x}+v_{0,\perp}\vec{e}_{y}.
\label{ass_mom_ref}
\end{equation}
For both signal and reference trajectories to lead to the same final momentum $\vec{p}=\left(p_x, p_y\right)$, their velocities at any time $t>t_{1}$ must be equal Ref.~\cite{Bian2011}. Indeed, from Eqs.~(\ref{vel_after_sig}) and (\ref{ass_mom_sig}) it follows that for $t>t_{1}$
\begin{equation}
\vec{V}\left(t\right)=\left[p_x-p_{F}\sin\left(\omega t\right)\right]\vec{e}_x+p_{y}\vec{e}_y.
\label{vel_eq}
\end{equation} 
The same expression is also valid for the reference electron [cf. Eqs.~(\ref{velocity}) and (\ref{ass_mom_ref})]. 

When the three-step model is applied to the SFPH, the phase associated with an electron trajectory starting at time $t_0$ is determined by the classical action (see Refs.~\cite{Bian2011,Bian2012, Bian2014, Li2014, Li2015}) $S\left(t_0,t\right)~=~\int_{t_0}^{t}\left(\vec{v}^{2}\left(t'\right)/2+I_{p}\right)dt'$, where $I_p$ is the ionization potential. Therefore, the phases of the signal and reference electrons are given by:
\begin{equation}
S^{sig}\left(t\right)=\frac{1}{2}\int_{t_{0}^{sig}}^{t_1}v_{x}^{2}\left(t'\right)dt'+\frac{1}{2}\int_{t_1}^{t}\left[p_x-p_{F}\sin\left(\omega t'\right)\right]^2dt'+I_{p}\left(t-t_{0}^{sig}\right)+\frac{p_{y}^{2}}{2}\left(t-t_1\right),
\label{phase_sig}
\end{equation}
and
\begin{equation}
S^{ref}\left(t\right)=\frac{1}{2}\int_{t_{0}^{ref}}^{t_1}v_{x}^{2}\left(t'\right)dt'+\frac{1}{2}\int_{t_1}^{t}\left[p_x-p_{F}\sin\left(\omega t'\right)\right]^2dt'+I_{p}\left(t-t_{0}^{ref}\right)+\frac{p_{y}^{2}}{2}\left(t-t_{0}^{ref}\right),
\label{phase_ref}
\end{equation}
respectively. Finally, the phase difference between the signal and reference waves reads as (see Refs.~\cite{Bian2011, Bian2012, Bian2014, Li2014, Li2015}):
\begin{equation}
\Delta S=\frac{1}{2}\int_{t_{0}^{sig}}^{t_1}v_{x}^{2}\left(t'\right)dt'-\frac{1}{2}\int_{t_{0}^{ref}}^{t_1}v_{x}^{2}\left(t'\right)dt'-I_{p}\left(t_0^{sig}-t_{0}^{ref}\right)-\frac{p_y^2}{2}\left(t_1-t_{0}^{ref}\right).
\label{phase_diff}
\end{equation}

In order to calculate the phase difference (\ref{phase_diff}) at a given final momentum $\vec{p}$, it is necessary to find the corresponding values of $t_0^{ref}$, $t_0^{sig}$, and $t_1$. For $\left|p_x\right|<p_{F}$ the equation $p_x~=~p_{F}\sin\left(\omega t_0^{ref}\right)$ [see Eq.~(\ref{ass_mom_ref})] has two solutions in the range $0\leq \omega t_0^{ref}< 2\pi$. For $p_x\geq0$ we have $\omega t_{0,1}^{ref}=\arcsin\left(p_x/p_{F}\right)$ and $\omega t_{0,2}^{ref}=\pi-\arcsin\left(p_x/p_{F}\right)$. For negative $p_x$ the solutions are given by $\omega t_{0,1}^{ref}=\pi+\arcsin\left(\left|p_x\right|/p_{F}\right)$ and $\omega t_{0,2}^{ref}=2\pi-\arcsin\left(\left|p_x\right|/p_{F}\right)$. 

Retrieving the corresponding $t_0^{sig}$ and $t_1$ is more cumbersome. It is worthwhile to solve Eq.~(\ref{retcond}) first, i.e., to find the function $t_1=t_1\left(t_0^{sig}\right)$ at every point of some grid for the time of start $t_0^{sig}$. For some values of ionization time Eq.~(\ref{retcond}) has multiple solutions, which correspond to so-called late returns of ionized electron to the ion. Knowing $t_1=t_1\left(t_0^{sig}\right)$ on some grid we can calculate $V_1\left(t_0^{sig}\right)$ on the same grid [see Eq.~(\ref{vel_up_ret})]. Then the values of $t_1$ and $V_1$ at any intermediate point can be found by interpolation. From Eqs.~(\ref{vel_up_ret}) and (\ref{vel_eq}) it follows that 
\begin{equation}
\left(p_x-p_{F}\sin\left[\omega t_1\left(t_0^{sig}\right)\right]\right)^2+p_{y}^{2}=p_{F}^{2}\left\{\sin\left[\omega t_1\left(t_0^{sig}\right)\right]-\sin\left[\omega t_0^{sig}\right]\right\}^2.
\end{equation} 
By solving this equation numerically we can find the time of start $t_0^{sig}$ that leads to a given $\vec{p}$, and then evaluate  the corresponding values of $t_1\left(t_0^{sig}\right)$ and $V_1\left(t_0^{sig},t_1\right)$. Finally, Eqs.~(\ref{ass_mom_sig}) allows us to determine the instantaneous scattering angle.

The algorithm described here is not the only possible approach to calculation of interferometric structures. An alternative  method treats $t_0^{sig}$ and $\theta_{0}$ as new independent variables. For every time of start $t_0^{sig}$ the corresponding recollision time is again found from Eq.~(\ref{retcond}). Then for any pair $\left(t_0^{sig},\theta_0\right)$ the asymptotic momentum can be found from Eq.~(\ref{ass_mom_sig}). By doing so for sufficiently large number of trajectories specified by $t_0^{sig}$ and $\theta_0$, we can obtain reliable statistics in the $\left(p_x,p_y\right)$ plane. Finally, these signal trajectories with the corresponding phases are binned according to their asymptotic momenta in cells in momentum space. After the phase of the reference electron leading to a given cell is found, we can calculate the phase difference $\Delta S$ associated with this cell. 

Instead of dividing the $\left(p_x,p_y\right)$ plane into cells, interpolation on the nonuniform grid can be applied to find the value of $S^{sig}$ for any given momentum $\vec{p}$. However, for the laser-atom parameters used in this work, the results of such interpolation converge slowly with increasing number of sets $\left(t_0^{sig},\theta_0\right)$. Moreover, before convergence is achieved, the interferometric structures calculated within this approach show some spurious structures, which could be confused with the true interference patterns. For this reason, we use bins on the $\left(p_x,p_y\right)$ plane when implementing the second approach with $t_0^{sig}$ and $\theta_0$ being the independent variables. We have used both approaches to check the consistency. The results obtained within both methods are in agreement. 

Four major different types of interference structures are usually discussed (see Ref.~\cite{Bian2011}). Two of these types correspond to forward scattering of a signal electron, whereas two other types involve its backward scattering. Nevertheless, in all these four types the signal electron starts from the first quarter of the laser period $0^{\circ}\leq\omega t_0\leq 90^{\circ}$. Therefore, we start our analysis with the kinematics of the signal electron launched in the first quarter of cycle.  

\subsection{Interference structures for time-independent exit point}

Let us first consider the simplest case that is usually assumed when the three-step model is used to calculate holographic interference patterns~\cite{Bian2011, Bian2012, Bian2014, Li2014}: the tunnel exit is equal for all ionization times and it is determined by the amplitude of the laser field: $x_0=I_p/F_0$. Figure 1 shows the asymptotic momentum components of such a signal electron as functions of the start time and instantaneous scattering angle for ionization of the H atom at a wavelength of 800 nm and two laser intensities: $2\times10^{14}$ W/cm$^2$ and $6\times10^{14}$ W/cm$^2$. For most values of the scattering angle $\theta_{0}$, the $p_x$-component of the asymptotic momentum has a minimum as a function of $\omega t_0$. This minimum (maximum of the absolute value) is particularly pronounced for the backward scattered electrons: $90^{\circ} \leq \theta_0 \leq 270^{\circ}$. Nevertheless, it may also exist for forward scattered electrons, i.e., for $0^{\circ} \leq \theta_0<90^{\circ}$ (or $270^{\circ}<\theta_0 \leq 360^{\circ}$). The presence of this minimum implies that for some values of the final momentum $\vec{p}$ there are two different ionization times corresponding to this angle $\theta_{0}^{*}$ [see Fig.~2 where the cuts of Figs.~1~(a) and (b) at $\theta_0=180^{\circ}$ are shown]. These two ionization times correspond to different electron trajectories. Depending on whether the signal electron is ionized before the minimum or after it, we refer to the corresponding trajectory as long or short one. Therefore, it is necessary to distinguish the interference structures created by reference and long signal trajectories from those produced by reference and short signal trajectories. This issue was also discussed in Ref.~\cite{Li2014}. 

\begin{figure}[h]
\begin{center}
\includegraphics[width=0.85\textwidth]{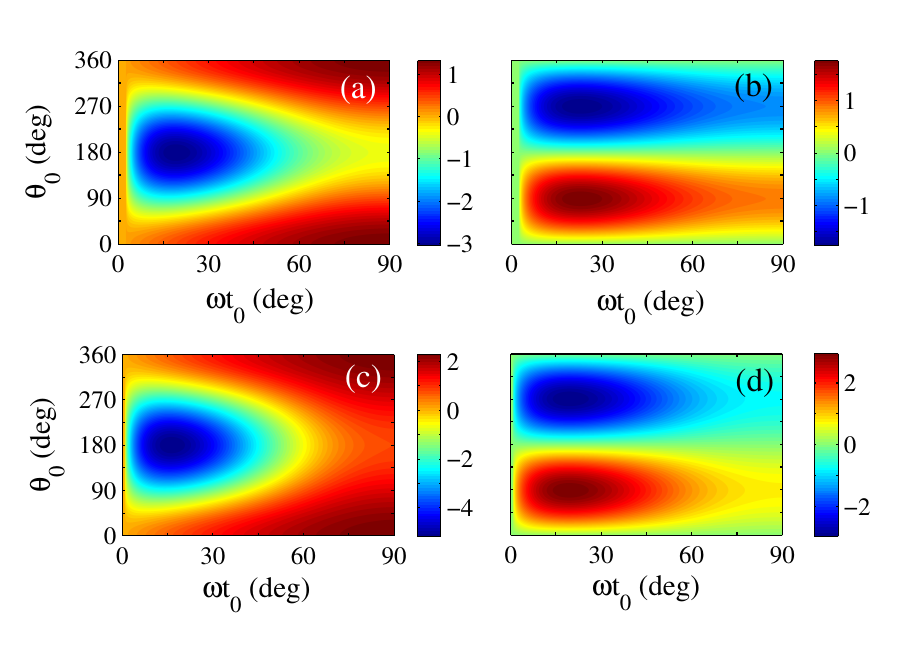} 
\end{center}
\caption{Asymptotic electron momentum components as functions of the ionization time and the instantaneous scattering angle for the H atom ionized by the first quarter of a cosine-like field with a wavelength of 800 nm (Ti:sapphire laser) polarized along the $x$-axis. The left column, that is, panels (a) and (c), shows the $p_x$-component. The right column [panels (b) and (d)] shows the $p_y$-component. Panels [(a),(b)] and [(c),(d)] correspond to the intensities of 2.0 and $6.0\times10^{14}$ W/cm$^2$, respectively.}
\label{fig1}
\end{figure} 

\begin{figure}[h]
\begin{center}
\includegraphics[width=0.70\textwidth]{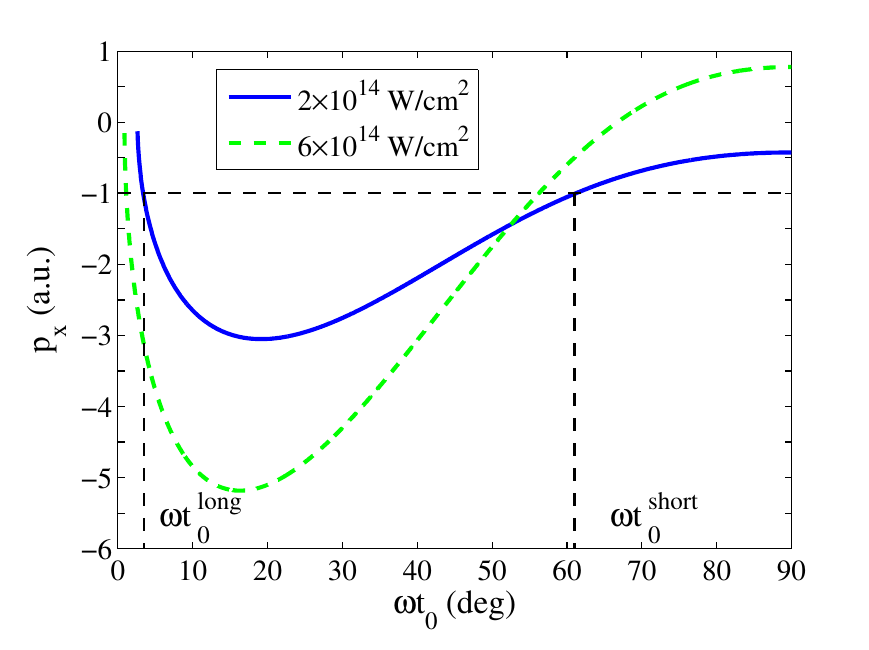} 
\end{center}
\caption{The $p_x$-component of the asymptotic electron momentum as a function of ionization time for the scattering angle $\theta_0=180^{\circ}$ (backward scattering). The parameters are as in Fig.~1. The blue (solid) and green (dashed) curves correspond to the intensities of 2.0 and $6.0\times10^{14}$ W/cm$^2$, respectively.}
\label{fig2}
\end{figure} 

The first and widely discussed type of the holographic interference emerges when both reference and signal electrons are launched in the first quarter of the optical cycle, see Fig.~3~(a). For brevity, we refer to this kind of interference as type A. Usually it is also assumed that the signal electron is scattered forward. The interference pattern of the second kind [type B, Fig.~3~(b)] is produced when the signal electron is launched on the first quarter cycle, whereas the reference electron is generated in the second quarter. The interference structures [i.e., $\cos\left(\Delta S\right)$, where $\Delta S$ is given by Eq.~(\ref{phase_diff})] of types A and B are shown in Figs.~3~(c) and (d). For the cosine-like field discussed here they emerge in the half plane $p_x>0$. In contrast to Refs.~\cite{Bian2011, Bian2012, Bian2014, Li2014}, we do not restrict our consideration to forward scattered electrons only. Instead, we allow for all instantaneous scattering angles $\theta_0$. The short and long signal trajectories should be separated in calculations. 
The averaging over the phase differences that correspond to the long and short trajectories ending up to a given $\vec{p}$ leads to the results shown in Figs.~3~(e) and (f): the false substructures are clearly visible on the edges of the interference patterns. In contrast to this, Figs.~3~(c) and (d) were obtained by considering only the long trajectories for the momenta at the edges of the structure. 

\begin{figure}[h]
\begin{center}
\includegraphics[width=0.70\textwidth]{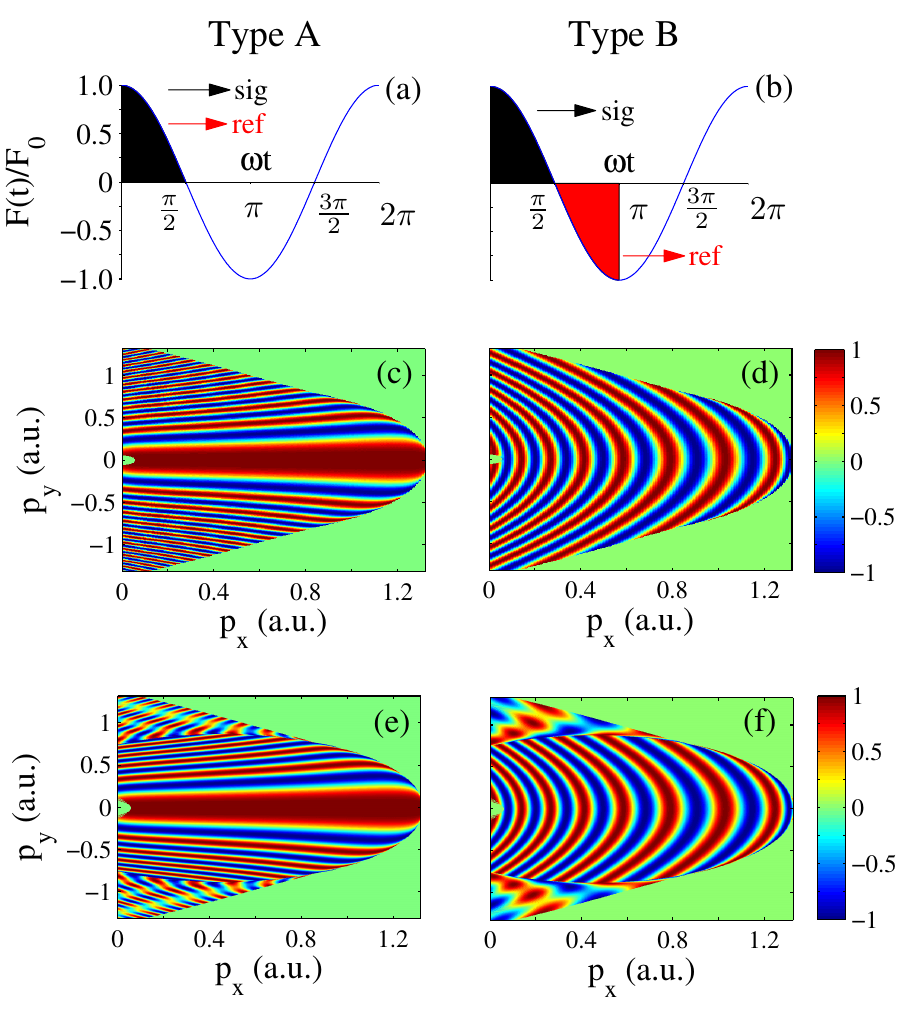} 
\end{center}
\caption{Forward scattering holographic interference patterns calculated within the three-step model for ionization of H atom by the laser field with a wavelength of 800~nm and intensity of $2.0\times10^{14}$~W/cm$^2$. Panels (a) and (b) illustrate formation of the interferences of types A and B, respectively. Panels (c) and (d) present the interference structure of types A and B, respectively. The same structures calculated without separation of different forward scattered signal trajectories leading to same final momenta are shown in panels (e) and (f), respectively.}
\label{fig3}
\end{figure} 

When the reference electron is launched in the third quarter of cycle, we refer to the corresponding structure as interference type C. In the case when the reference trajectory starts in the fourth quarter, we classify the interference structure as type D. In order to distinguish between the patterns created by long and short signal trajectories, we add the word ``long'' or ``short''. Thus we consider the following types of holographic patterns: C-long, C-short, D-long, and D-short [see Figs.~4 (a), (b), (e), and (f), respectively]. The corresponding interference structures calculated at the intensity of $2.0\times 10^{14}$ W/cm$^2$ are shown in Figs.~4 (c), (d), (g), and (h). Note that, in contrast to Refs.~\cite{Bian2011, Bian2012, Bian2014, Li2014}, when calculating the structures of types C-long and D-long we do not restrict ourselves to backward scattered electrons only, i.e., we assume $0^{\circ}\leq \theta_0 \leq 360^{\circ}$. However, only the backward scattered electrons are taken into account in calculations of the patterns of types C-short and D-short. This is due to the fact that the interference of the short forward scattered electrons with the reference ones results in a different kind of interference structure emerging in a small part of the $\left(p_x, p_y\right)$ plane. This structure is outside of the scope of the present work.

\begin{figure}[h]
\begin{center}
\includegraphics[width=0.75\textwidth]{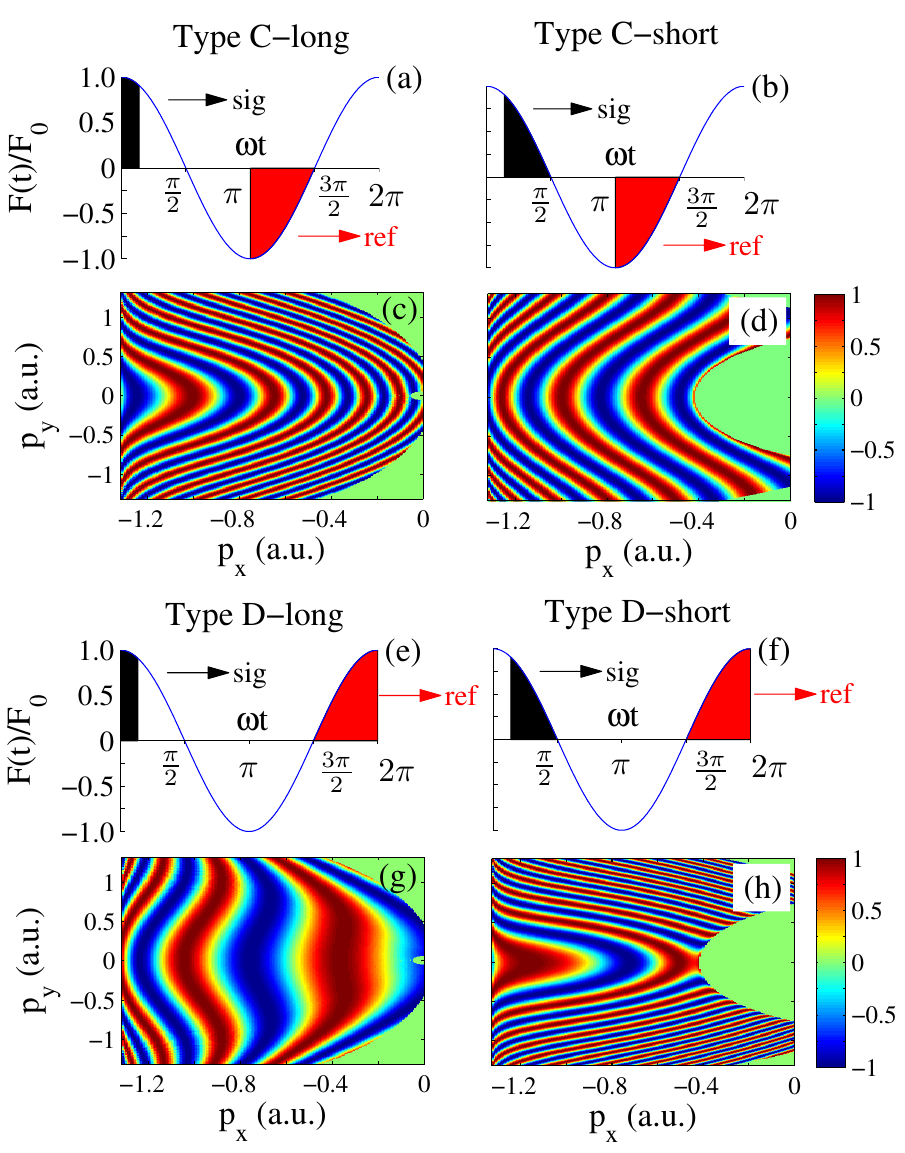} 
\end{center}
\caption{Backward scattering holographic interferences predicted by the three-step model. Panels (a), (b), (e), and (f) show schematic illustration of the structures of types C-long, C-short, D-long, and D-short, respectively. Panels (c), (d), (g), and (h) present the corresponding interference patterns. The parameters are as in Fig.~3.}
\label{fig4}
\end{figure} 

Special attention must be given to the area of the $\left(p_x, p_y\right)$ plane in Figs.~4~(d) and (h), where the interference structure is absent. The vanishing of the interference structure occurs due to the fact that not all the values of the $p_x$ component can be reached by short trajectories, see Figs.~1~(a) and (b). This effect is particularly pronounced at lower intensities and disappears with increasing field strength. Interference structures of all six types discussed here are shown in Fig.~5 at the higher intensity $6.0\times10^{14}$ W/cm$^2$. With increasing intensity the interference stripes become narrower and their number increases dramatically. At this higher intensity the holographic structures of types C-short and D-short fill the half of the plane $-F_0/ \omega < p_x <0$ completely.   

\begin{figure}[h]
\begin{center}
\includegraphics[width=0.75\textwidth]{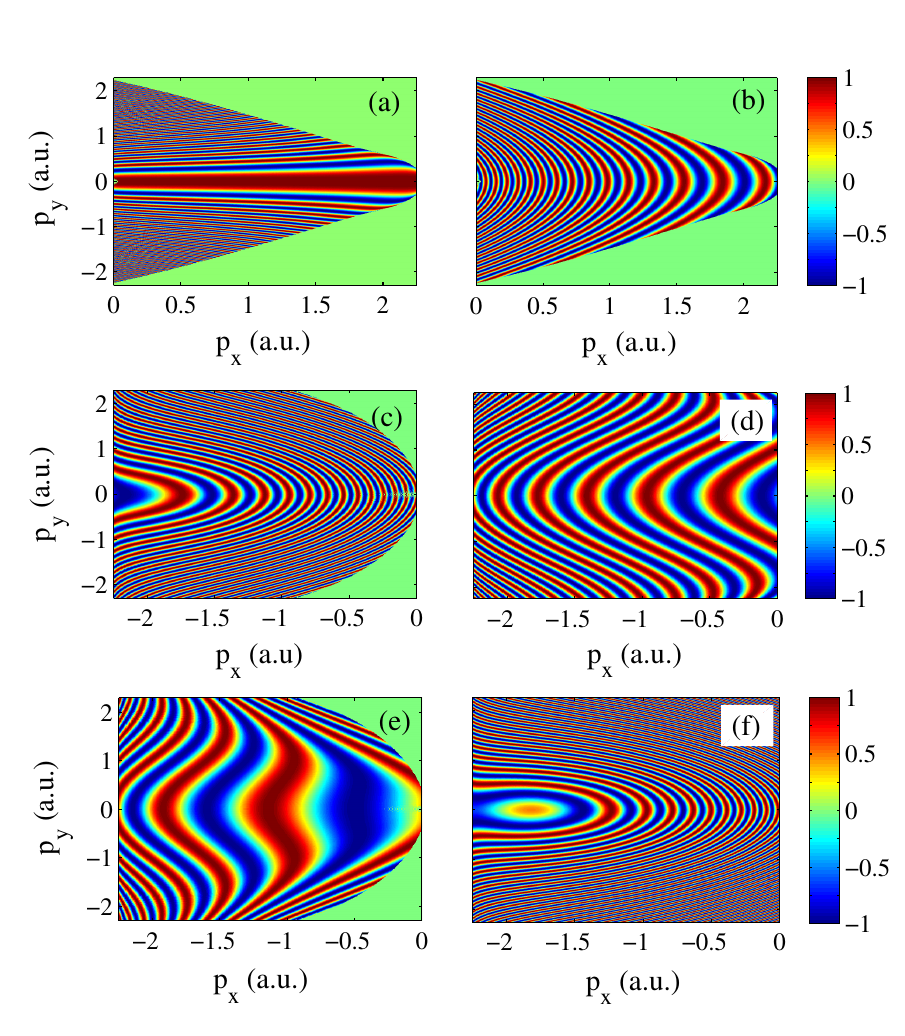} 
\end{center}
\caption{Holographic interference patterns obtained from the three-step model for ionization of H by a Ti:sapphire laser (800 nm) at the intensity $6.0\times10^{14}$~W/cm$^2$. Panels (a), (b), (c), (d), (e), and (f) show the interference structures of types A, B, C-long, C-short, D-long, and D-short, respectively.}
\label{fig5}
\end{figure} 

\subsection{Interferometric structures for time-dependent exit point}

Next, we analyze the interference structures calculated assuming that the starting point of the trajectory depends on time: $x_0\left(t\right)=-I_{p}/F_{0}\left(t\right)$. First, we recalculate $p_x$ and $p_y$ components as functions of $\omega t_0^{sig}$ and $\theta_0$ [see Fig.~6 (a) and (b)]. The results differ dramatically from the case of constant $x_0$ [cf. Figs.~1 (c) and (d)]. Indeed, the function $p_x=p_x\left(\omega t_0, \theta_0\right)$ has now two minima. Accordingly, the $p_y$ component has two maxima for $0^{\circ} \leq \theta_0 \leq 180^{\circ}$ and two minima for $180^{\circ} \leq \theta_0 \leq 360^{\circ}$. The second extremum of $p_x$ and $p_y$ in the vicinity of $\omega t_0=90^{\circ}$ gives rise to another kind of interference structures. Here we do not consider this second extremum and, therefore, account only for the electrons launched at $\omega t_0\leq67^{\circ}$. 

\begin{figure}[h]
\begin{center}
\includegraphics[width=0.85\textwidth]{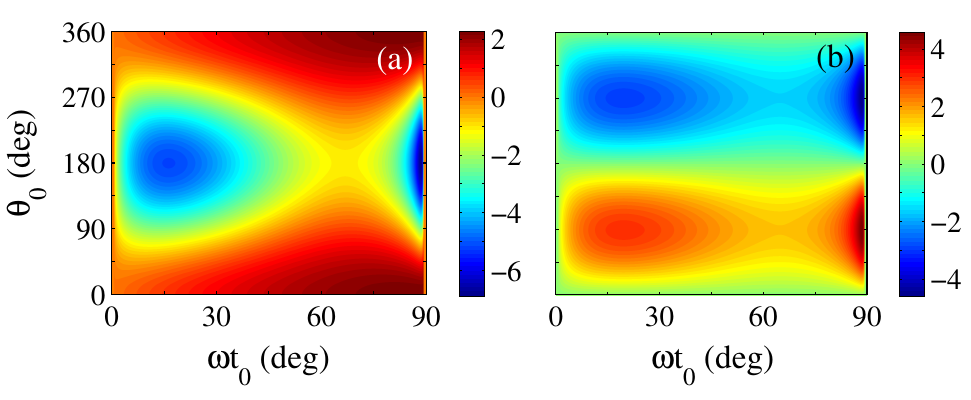} 
\end{center}
\caption{Same as Fig.~1 for time-dependent exit point and intensity $6.0\times10^{14}$ W/cm$^2$.}
\label{fig6}
\end{figure} 

The corresponding interference patterns are shown in Figs.~7~(a)-(f). It is seen that the structures of types A and B occupy larger area on the $\left(p_x,p_y\right)$ plane compared to those shown in Figs.~5~(a) and (b). The caustics that are hardly visible in Figs.~7(a) and better seen in Fig.~7~(b) arise due to the discontinuity of the $\omega t_0\left(p_x, p_y\right)$ found from Eq.~(\ref{retcond}) for $x_0$ depending on $t_0$. The time-dependent exit point leads to the decrease of the area occupied by the interference structures of types C-short and D-short, see Fig.~7~(d) and (f). In this respect, the account for the time dependence in the expression for the exit point has a similar effect as the decrease of the laser intensity. This result is expected, because the maximum value $F_0$ of the laser field $F\left(t\right)$ was used when calculating Fig.~1 with fixed $x_0$.  

\begin{figure}[h]
\begin{center}
\includegraphics[width=0.75\textwidth]{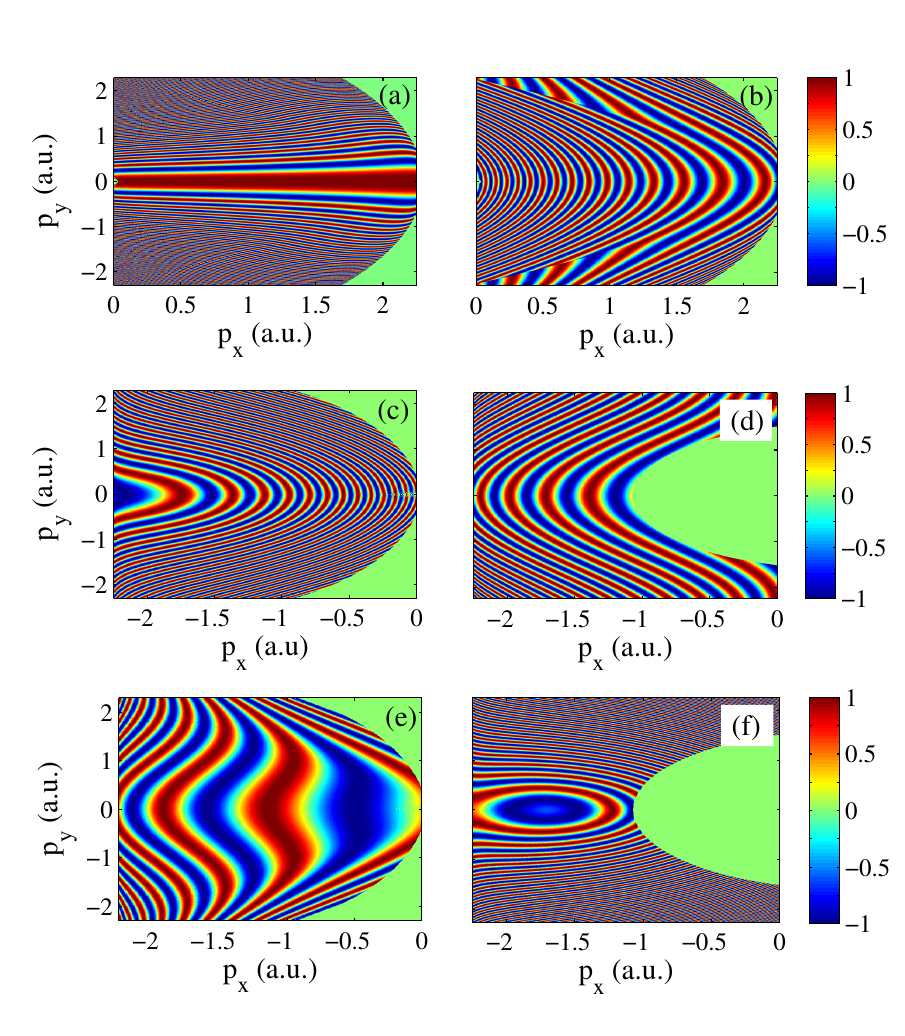} 
\end{center}
\caption{Same as Fig.~5 for time-dependent exit point.}
\label{fig7}
\end{figure} 



\section{Calculation of the holographic interference patterns with Coulomb potential}

In order to calculate the holographic interference patterns in the presence of the Coulomb field, we use an adapted version of the SCTS model. Here we sketch the main points of our approach focusing on the differences in the implementation compared to Ref.~\cite{Shvetsov2016}. As in any semiclassical approach, the trajectory $\vec{r}\left(t\right)$ and momentum $\vec{p}\left(t\right)$ of an electron in the SCTS model are calculated using Newton's equation of motion:
\begin{equation}
\frac{d^2\vec{r}}{dt^2}=-\vec{F}\left(t\right)-\frac{Z\vec{r}\left(t\right)}{r^3\left(t\right)},
\label{newton}
\end{equation} 
where $Z$ is the ionic charge ($Z=1$ for the H atom). In our simulations we solve Eq.~(\ref{newton}) for the electrons launched on the central period of the Ti:sapphire laser pulse (800 nm) with the full duration of 8 optical cycles. In order to integrate Eq.~(\ref{newton}), we need to specify initial conditions, i.e., the initial velocity and position. To this end, one could use the separation of the static tunneling problem in parabolic coordinates \cite{Dau3, Bisgaard2004, Dimitrovski2010}. However, in the present work we use the simplest formula for the tunnel exit neglecting the Coulomb potential: $x_0=-I_p/F\left(t_0\right)$ to allow for direct comparison with the results of the three-step model. 

We assume that the electron starts with zero initial velocity along the laser polarization $v_x\left(t_0\right)=0$ and nonzero initial velocity $v_{0,\perp}$ in the transverse direction. The SCTS considers an ensemble of classical trajectories with different $t_0$ and $v_{0,\perp}$. Since in this work we are interested in the holographic interference structures rather than in calculation of the momentum distributions, we disregard trajectory weights and distribute the trajectories uniformly.

Following the SCTS model we associate every trajectory with the phase of the semiclassical propagator (see Refs.~\cite{Miller1971, Walser2003, Spanner2003}). For the Coulomb potential this phase is given by
\begin{equation}
\Phi\left(t_0,\vec{v}_{0}\right)=-\vec{v}_{0}\cdot\vec{r}\left(t_0\right)+I_pt_0-\int_{t_0}^{\infty}dt\left\{\frac{p^2\left(t\right)}{2}-\frac{2Z}{r\left(t\right)}\right\},
\label{que}
\end{equation} 
see Ref.~\cite{Shvetsov2016}. Once the asymptotic momenta of all the trajectories in the ensemble are found, we bin them in cells in momentum space. Finally, the amplitudes $\exp\left[i\Phi\left(t_0^{j},\vec{v}_0^{j}\right)\right]$ with $j=1,...,n_p$ associated with all $n_p$ trajectories ending up in a given bin located at $\vec{p}=\left(p_x,p_y\right)$ can be added coherently. However, it is easy to see that the quantity
\begin{equation}
Q=\left|\sum_{j=1}^{n_p}\exp\left[i\Phi\left(t_0^{j},\vec{v}_0^{j}\right)\right]\right|^2,
\label{qdef}
\end{equation}
which is similar to the ionization probability calculated according to the importance sampling implementation of the SCTS model, is not sufficient to obtain the phase difference between signal and reference electrons we are interested in. Indeed, for calculation of interference patterns similar to those shown in Figs.~3,4,5, and 7, we have to isolate \textit{only one} kind of rescattered and \textit{only one} kind of direct trajectories. In the presence of the Coulomb field this is not an easy task. Nevertheless, this objective can be accomplished by careful choice of initial conditions, i.e., of $t_0$ and $v_{0,\perp}$. 

Once the necessary isolation of different kinds of trajectories is achieved, in any bin of the momentum space we have $N$ trajectories of one kind with phases $\Phi_{0}^{i}$ $\left(i=1,...,N\right)$ and $M$ trajectories of another kind with phases $\Phi_{1}^{k}$ $\left(k=1,...,M\right)$. Then we calculate the average cosine of the phase difference $\left\langle \cos\left(\Phi_{0}-\Phi_{1}\right)\right\rangle$
in every bin of the momentum plane. 

\section{Strong-field photoelectron holography with the Coulomb potential}

We begin our analysis of effects of the Coulomb potential with the intracycle interference, i.e., the interference of reference (direct) electrons starting from two different quarters of the laser period, see, e.g., Refs.~\cite{Gopal2009, Arbo2010, Bian2011, Faria2017b}. The corresponding interference patterns produced by the reference electrons launched on the first and on the second quarter of the period calculated within the three-step model and with the account for the Coulomb field are shown in Figs.~8~(a) and (b), respectively. It is apparent that the Coulomb field creates characteristic kinks in the vicinity of $p_y=0$. The interference pattern of Fig.~8~(b) is similar to the structures that are seen in the momentum distributions calculated in Ref.~\cite{Faria2017b}.  

\begin{figure}[h]
\begin{center}
\includegraphics[width=0.90\textwidth]{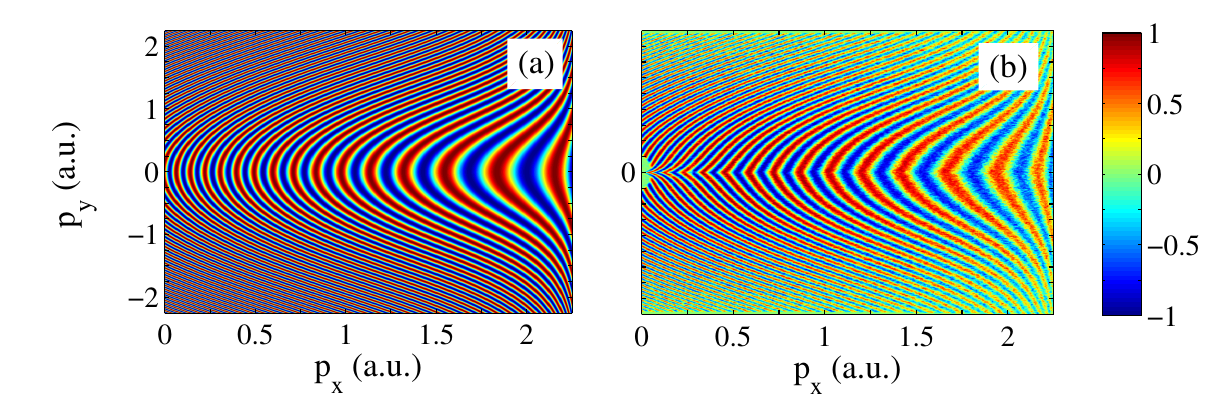} 
\end{center}
\caption{Interference pattern for intracycle interference obtained (a) from the three-step model and (b) in the presence of the Coulomb potential. The parameters are as in Fig.~5.}
\label{fig8}
\end{figure} 

To understand the formation of the interference patterns in the presence of the Coulomb field, one must note that the reference trajectories are considered to be those that pass the core at large distances and undergo small-angle scattering, whereas the signal trajectories are those that pass the parent ion at small distances and undergo large-angle scattering causing a sign change of the momentum in $y$-direction. Therefore, it appears natural to consider only the electrons obeying the condition $v_{0,\perp}p_{y}\geq0$ when calculating of Fig.~8~(b). However, this restriction is not sufficient to calculate interference patterns properly. The reason is the non-trivial dependence of the final electron momentum on the initial conditions in the presence of the Coulomb field. Indeed, the calculation of the $p_y$-component as a function of $t_0$ and $v_{0,\perp}$ shows that for each time of start $t_0$ there is a smallest positive initial transverse velocity $v_{0,\perp}^{+}\left(t_0\right)$ [largest negative $v_{0,\perp}^{-}\left(t_0\right)$]  leading to $p_y\geq0$ [$p_y\leq0$]. This implies that $p_y\geq0$ [$p_y\leq0$] for any $v_{0,\perp}>v_{0,\perp}^{+}\left(t_0\right)$ [$v_{0,\perp}<v_{0,\perp}^{+}\left(t_0\right)$]. Moreover, for certain ranges of ionization time $t_0$, the $y$-component of the final momentum is not a sign-constant function over the intervals $v_{0,\perp}^{-}\left(t_0\right)<v_{0,\perp}\leq v_{0,\perp}^{+}\left(t_0\right)$. As a result, some trajectories launched with $0<v_{0,\perp}\leq v_{0,\perp}^{+}\left(t_0\right)$ [$0>v_{0,\perp}\geq v_{0,\perp}^{-}\left(t_0\right)$] are detected with $p_y\geq 0$ [$p_y<0$]. These trajectories interfere with those starting from another quarter of cycle with $v_{0,\perp}> v_{0,\perp}^{+}\left(t_0\right)$ [$v_{0,\perp}<v_{0,\perp}^{-}\left(t_0\right)$], and, therefore, create an additional interference pattern in some part of the $\left(p_x, p_y\right)$ plane. This pattern should not be mixed with the main one created solely by the electrons with $v_{0,\perp}> v_{0,\perp}^{+}\left(t_0\right)$ [$v_{0,\perp}<v_{0,\perp}^{-}\left(t_0\right)$]. Accordingly, from this point on we exclude the reference trajectories not obeying the condition $v_{0,\perp}> v_{0,\perp}^{+}\left(t_0\right)$ [$v_{0,\perp}<v_{0,\perp}^{-}\left(t_0\right)$]. 

Another significant point is that the first quarter of the laser cycle alone is no longer sufficient to produce the whole interference pattern when the Coulomb field is taken into account. The interference structure shown in Fig.~8~(b) was calculated by considering the range of ionization times $\omega t_0$ between $-10^{\circ}$ and $90^{\circ}$. This shift of the left boundary compared to the case of the three-step model is caused by the change of the final electron momentum due to the Coulomb potential. The magnitude of this shift can be easily estimated by treating the Coulomb field as a perturbation. Its contribution to the asymptotic momentum is calculated by integrating the Coulomb force along the trajectory governed by the laser field only ~\cite{Shvetsov2004, Shvetsov2009}. For not very large transverse velocities $v_{0,\perp}<\sqrt{2I_p}$ the corresponding integral can be evaluated analytically. As a result, for the cosine-like field $\vec{F}\left(t\right)=F_{0}\cos\left(\omega t\right)\vec{e}_x$ the $x$-component of the final momentum $\vec{p}$ can be estimated as (see Ref~\cite{Shvetsov2009} for details): 
\begin{equation}
\label{est1}
p_x \approx \frac{F_0}{\omega}\sin\left(\omega t_0\right)+\pi \frac{F_0\cos\left(\omega t_0\right)}{\left(2I_p\right)^{3/2}} 
\end{equation}  
It is easy to see that Eq.~(\ref{est1}) can be rewritten as 
\begin{equation}
p_x \approx F_0\sqrt{\frac{1}{\omega^2}+\frac{\pi^2}{\left(2I_p\right)^3}}\cos\left(\omega t_0 - \alpha\right),
\label{trig}
\end{equation}
where
\begin{equation}
\alpha=\arctan\left[\frac{\left(2I_p\right)^{3/2}}{\omega\pi}\right].
\label{alp}
\end{equation}
The left edge of the intracycle interference structure in the three-step model is located at $p_x=0$ [see Fig.~3~(a)], corresponding to $\omega t_{0}=0$. In the presence of the Coulomb potential, it follows from Eq.~(\ref{trig}) that for $p_x$ to be equal to zero, $\omega t_0=\alpha+\frac{\pi}{2}+\pi k$, where $k$ is an integer number. Therefore, the ionization time closest to $\omega t_0=0$ that leads to $p_x=0$ is estimated as:  
\begin{equation}
\omega t_0^{\star}\approx -\frac{\pi}{2}+\arctan\left[\frac{\left(2I_p\right)^{3/2}}{\omega\pi}\right]=-\arccot\left[\frac{\left(2I_p\right)^{3/2}}{\omega\pi}\right]. 
\label{deg10}
\end{equation} 
For the parameters of Fig.~8 this estimate yields $\omega t_0^{\star}=-10.2^{\circ}$. Note that Eq.~(\ref{deg10}) does not depend on the field strength. More accurate expressions for the asymptotic momentum of a direct electron moving in laser and Coulomb fields were obtained recently ~\cite{Becker2014, Kelvich2016}. Furthermore, analytical estimates for the final momenta of different kinds of rescattered electrons were also derived in Refs.~\cite{Becker2014, Kelvich2016}. However, the simple formula (\ref{est1}) is sufficient to understand the formation of the holographic interference patterns. Note that the vast majority of the reference trajectories starting in the vicinity of $\omega t_{0}=\frac{\pi}{2}$ are only weakly affected by the Coulomb field. For this reason, we do not shift the right boundary of the first quarter (the left boundary of the second quarter) of cycle. 

In the three-step model, the interference structure of type A corresponds to the situation when both signal and reference electrons are launched within the first quarter of the laser period. In contrast to type A, the structure of type B is created by the signal electrons launched on the first quarter of cycle and the reference trajectories starting on the second quarter. When the Coulomb field is taken into account, the signal trajectories must obey the condition $v_{0,\perp}p_{y}<0$. Furthermore, the $x$-component of the final momentum must be positive, since the interference pattern of type A emerges in the half-plane $p_x>0$. However, these conditions are not sufficient to produce proper interference pattern (similarly to the case of the intracycle interference). The reason is that the mapping from the $\left(t_0,v_{0,\perp}\right)$ plane to the $\left(p_x,p_y\right)$ plane is not a one-to-one function in the domain where the condition for the signal trajectories $v_{0,\perp}p_{y}<0$ is fulfilled. Different initial conditions $\left(t_0,v_{0,\perp}\right)$ can lead to the same final momentum $\vec{p}$. As the result, several different interference patterns can emerge in the same area of the $\left(p_x, p_y\right)$ plane. These patterns must be separated.

To this end, we identify the domains of initial conditions, where the mapping \linebreak ${\left(t_0,v_{0,\perp}\right)\rightarrow\left(p_x, p_y\right)}$ is a one-to-one function. As a result, we find that interference patterns similar to those obtained within the three-step model also emerge when the Coulomb field is taken into account. Figures 9 (a) and (b) show the $x$- and $y$-components of the final electron momentum as functions of $t_0$ and $v_{0,\perp}\geq 0$, respectively. The domain of the $\left(t_0,v_{0,\perp}\right)$ plane that gives rise to the relevant interference patterns similar to the one predicted by the three-step model is shown by the blue (dashed) curves in Figs.~9~(a) and (b). The corresponding interference patterns of types A and B at the intensity of $6.0\times10^{14}$ W/cm$^2$ are shown Figs.~10~(a) and (b). It is apparent that the Coulomb potential changes the interference structure substantially. This is true for the positions of the interference maxima and minima, as well as for their spacing. For type B the interference stripes show kinks at $p_y=0$, similar as in intracycle interference, Fig.~8~(b), but pointing in the opposite direction. 

\begin{figure}[h]
\begin{center}
\includegraphics[width=0.90\textwidth]{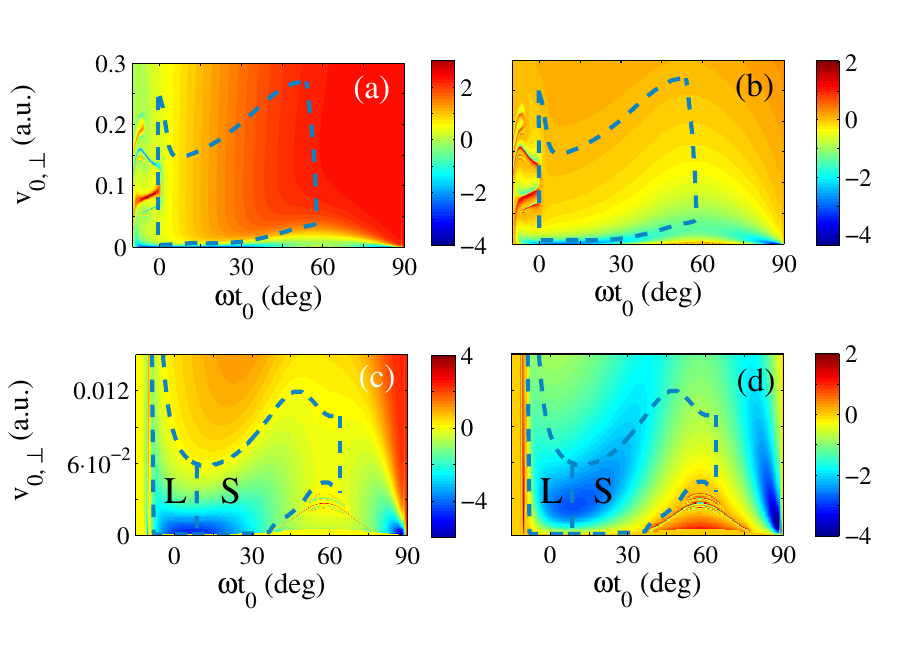} 
\end{center}
\caption{Asymptotic electron momentum components as functions of the ionization time and the initial transverse velocity calculated in the presence of the Coulomb potential. The parameters are as in Fig.~5. Panels (a) and (c) show the $p_x$-component. Panels (b) and (d) display the $p_y$-component. The blue (dashed) curves in panels (a) and (b) show the boundary of the domain that gives rise to the signal electrons responsible for the lower half $\left(p_y<0\right)$ of the interference patterns of types A and B. The letters ``L'' and ``R'' in panels (c) and (d) mark the domains responsible for the long and short signal trajectories, respectively. The boundaries of these domains are indicated by blue (dashed) curves.}
\label{fig9}
\end{figure} 

\begin{figure}[h]
\begin{center}
\includegraphics[width=0.75\textwidth]{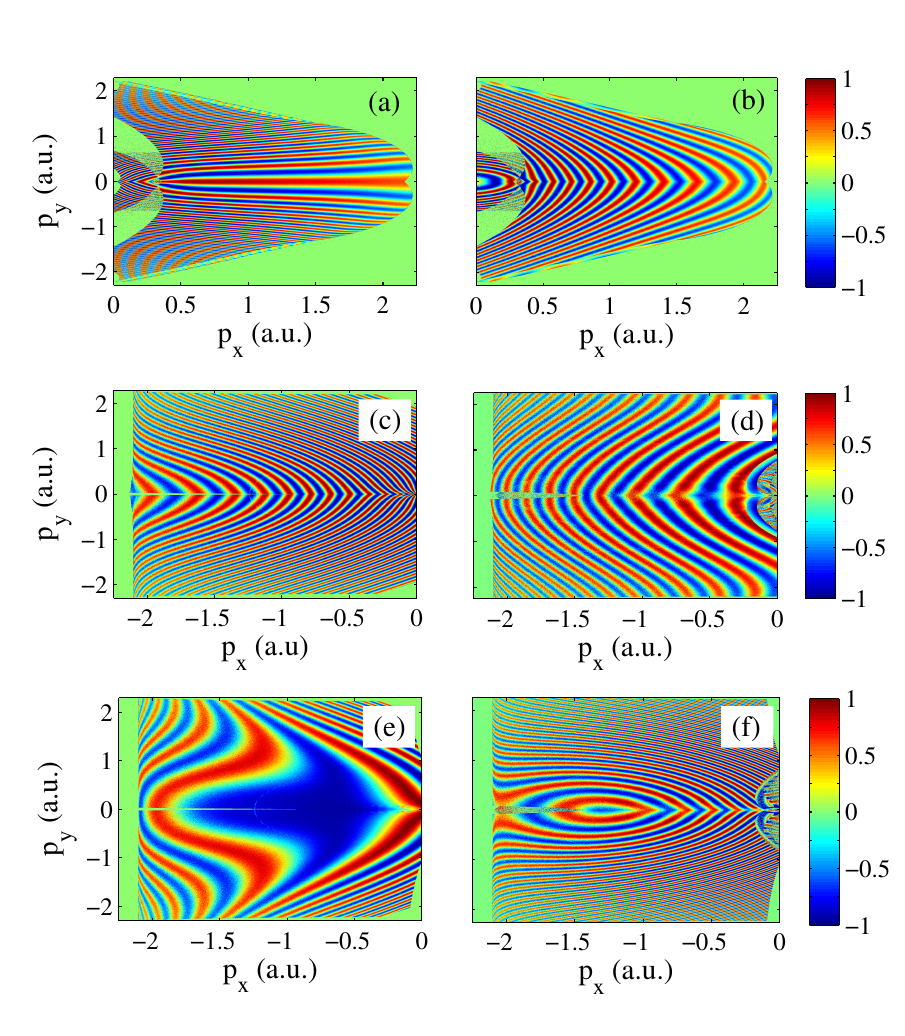} 
\end{center}
\caption{Holographic interference patterns of types A [panel (a)], B [panel (b)], C-long [panel (c)], C-short [panel (d)], D-long [panel (e)], and D-short [panel (f)] in the presence of the Coulomb potential. The parameters are as in Fig.~5.}
\label{fig11}
\end{figure}  

To obtain the interference structures of types C and D, we calculate the interference of the signal electrons with $v_{0,\perp}p_{y}<0$ starting from the first quarter with the reference electrons starting from the third or the fourth quarter of cycle. The interference patterns of types C and D correspond to negative $p_x$. 

The $x$- and $y$-components of the final electron momentum as functions of $\omega t_0\in\left[-10^{\circ},90^{\circ}\right]$ and small positive $v_{0,\perp}$ are shown in Figs.~9~(c) and (d). It is seen that both $p_x$ and $p_y$ components have two pronounced minima as in the case of the electron motion in the laser field only. The second minimum close to $\omega t_0=90^{\circ}$ is due to the time-dependent exit point [cf. Fig.~6]. Here we again consider only the left minimum of $p_x$ and $p_y$. The position of this minimum weakly depends on the initial transverse velocity. For the parameters considered the minimum is achieved at $\omega t_0\approx18.3^{\circ}$. Thus, by analogy with the electron kinematics within the three-step model, we can again distinguish between long and short signal trajectories. The corresponding domains on the $\left(t_0, v_{0,\perp}\right)$ plane are marked in Figs.~9~(c) and (d) by the letters ``L'' and ``R'', respectively. The boundaries of these domains are shown by blue (dashed) curves. With signal trajectories from these domains, we calculate interferences of types C-long, C-short, D-long, and D-short. As for the reference electrons, we proceed similarly to the cases of the intracycle interference and types A and B. Note that we shift the left boundary of the third quarter cycle by $10.2^{\circ}$ to earlier times when calculating the structure of type C. The results are presented in Figs.~10~(c), (d), (e), and (f). The patterns shown here should be compared with those presented in Figs~7~(c), (d), (e), and (f). It is seen that the Coulomb potential has three main effects on the interference patterns. First, it shifts the interference pattern as a whole. Second, it fills the parts of the interference structures that are absent when the Coulomb field is neglected. Third, the presence of the Coulomb potential results in the characteristic kinks of the interference stripes at $p_y=0$.  

Finally, we have checked the sensitivity of the results to changes in the exit point. To this end, we have recalculated the results shown using the expression for the tunnel exit that results from the separation of the time-independent Schr\"{o}dinger equation in parabolic coordinates. For the parameters under investigation, the corresponding interference structures are almost identical to those presented here.  

\section{Conclusions}

In conclusion, we have investigated subcycle interference structures emerging in strong-field photoelectron holography using semiclassical approaches. At first, we have calculated these structures within the three-step model. Following Refs.~\cite{Bian2011} we assumed initially that all classical electron trajectories start at the same point, which is determined as the tunnel exit at the maximum of the field. We found it important to distinguish between long and short rescattered trajectories when calculating the interference structures involving backscattering. This is in agreement with the conclusion of Ref.~\cite{Li2014}. 

We have found that the interference structures change significantly when the time dependence of the tunnel exit is taken into account. Specifically, some interference patterns expand, whereas others may shrink compared to those calculated with time-independent exit point. This is due to the substantial change in kinematics of the signal electron. 

In order to calculate the interference patterns in the presence of the Coulomb potential, we have developed a computational approach based on the SCTS model, which describes quantum interference including the Coulomb potential beyond the semiclassical perturbation theory. We have identified the specific groups of trajectories responsible for each kind of holographic structure. Finally, for every type of interference structure predicted by the three-step model we have presented its counterpart emerging in the presence of the Coulomb potential. In addition to changing the positions and the widths of the interference stripes, the Coulomb potential can manifest itself in three other effects. These are: the shift of the interference pattern as a whole, the filling of the parts of the interference structure that are missing when the Coulomb potential is neglected, and the characteristic kink of the interference stripes at zero transverse momentum. In measurable momentum distributions, several of the interference structures will usually be overlaid on top of each other. Therefore, future work is needed to shed light on the question which of the Coulomb effects are observable in momentum distributions.

\end{document}